\documentclass[aps, prd, floatfix, nofootinbib, twocolumn, superscriptaddress]{revtex4-2}
\usepackage{latexsym}
\usepackage{amsmath}
\usepackage{amssymb}
\usepackage{amsfonts}
\usepackage{textcomp}
\usepackage{color}
\usepackage{CJKutf8}

\usepackage[mathscr,scaled=1.15]{urwchancal}
\DeclareFontFamily{OT1}{pzc}{}
\DeclareFontShape{OT1}{pzc}{m}{it}%
{<-> s * [1.15] pzcmi7t}{}
\DeclareMathAlphabet{\mathpzc}{OT1}{pzc}{m}{it}

\usepackage{color}

\usepackage{supertabular}
\usepackage{placeins}
\usepackage{epsfig}
\usepackage{graphicx}

\definecolor{purple}{rgb}{0.5,0,0.5}
\definecolor{blue}{rgb}{0.0,0,0.9}
\definecolor{prdblue}{rgb}{0.133,0.118,0.498}
\usepackage[colorlinks=true, pdfstartview=FitV, linkcolor=prdblue, citecolor= prdblue, urlcolor=prdblue]{hyperref}

\usepackage{CJKutf8}

\begin{document}

\begin{CJK*}{UTF8}{gbsn}

\preprint{APS/123-QED}

\title{A Poincar\'e-covariant study of strange quark stars}

\author{\mbox{Hao-Ran Zhang (张浩然)}%
        $\,^{\href{https://orcid.org/0009-0005-3091-393X}{\textcolor[rgb]{0.00,1.00,0.00}{\sf ID}}}$}
\affiliation{School of Physics, Nanjing University, Nanjing, Jiangsu 210093, China}

\author{\mbox{Bo-Lin Li (李伯林)}%
        $\,^{\href{https://orcid.org/0000-0002-6348-604X}{\textcolor[rgb]{0.00,1.00,0.00}{\sf ID}}}$}
\email[]{blli@usst.edu.cn}
\affiliation{School of Physics, Faculty of Basic Sciences, University of Shanghai for Science and Technology, Shanghai 200093, China}

\author{\mbox{Zhu-Fang Cui (崔著钫)}%
        $\,^{\href{https://orcid.org/0000-0003-3890-0242}{\textcolor[rgb]{0.00,1.00,0.00}{\sf ID}}}$}
\email[]{phycui@nju.edu.cn}
\affiliation{School of Physics, Nanjing University, Nanjing, Jiangsu 210093, China}

\date{\today}

\begin{abstract}
We investigate the properties of dense quark matter and strange quark stars within a nonperturbative, Poincar\'e-covariant framework. Employing a symmetry-preserving vector$\,\otimes\,$vector contact interaction model, we extend the quark gap equation to the regime of zero temperature and finite quark chemical potential. From the resulting momentum-independent quark propagator, we construct the equation of state (EOS) and solve the Tolman-Oppenheimer-Volkoff (TOV) equations to evaluate the mass-radius relations and tidal deformabilities of strange quark stars. We systematically analyze the sensitivity of the EOS and the macroscopic stellar properties to the model parameters, specifically the effective interaction strength and the ultraviolet cutoff. We demonstrate that reducing the coupling constant stiffens the EOS, whereas increasing the ultraviolet cutoff softens it. By confronting our predictions with multi-messenger astrophysical constraints—including pulsar mass measurements and gravitational-wave data—we identify parameter regimes that successfully describe current observations. Specifically, we find that parameter sets with $\alpha_{ir}=0.735\pi$, $\Lambda_{uv}=0.905\,\mathrm{GeV}$ and $\alpha_{ir}=0.588\pi$, $\Lambda_{uv}=0.9955\,\mathrm{GeV}$, alongside a vacuum bag pressure of $B \approx (0.106\,\mathrm{GeV})^4$, yield stellar properties in excellent agreement with empirical constraints.

\end{abstract}

\maketitle

\end{CJK*}


\section{INTRODUCTION}\label{sec:introduction}

Compact stars provide a unique, natural laboratory for studying strongly interacting matter at extreme densities. Specifically, they offer access to the low-temperature, high-baryon-density regime of the quantum chromodynamics (QCD) phase diagram—a region currently inaccessible to terrestrial high-energy collider experiments \cite{Xu:2015vna,Arbanil:2016wud,Cui:2016zqp,Baym:2017whm,Li:2017zny}. In recent years, the era of multi-messenger astronomy, heralded by precise X-ray measurements from the NICER mission and gravitational-wave detections by the LIGO-Virgo-KAGRA collaborations, has yielded an unprecedented wealth of observational data on compact stars. 

Concurrently, the Bodmer-Witten hypothesis posits that strange quark matter — a phase comprising roughly equal numbers of up, down, and strange quarks—may constitute the true, absolute ground state of strongly interacting matter at zero pressure. According to this hypothesis, strange quark stars represent the stable ground state of compact stars, while conventional neutron stars are merely metastable or excited configurations \cite{Weber_2005,Yang:2023haz}. Therefore, strange quark stars serve as a vital testing ground for exploring the nonperturbative dynamics of dense QCD and the macroscopic manifestations of strong interactions.

Numerous theoretical models have been developed to investigate the internal structure and properties of compact stars, including chiral effective field theory (EFT)~\cite{Hebeler:2010jx,Hebeler:2015hla}, high-density perturbative QCD (pQCD) \cite{Kraemmer:2003gd,Fraga:2016yxs}, phenomenological bag models \cite{Alcock:1986hz}, and Nambu-Jona-Lasinio (NJL) type models \cite{Buballa:2003qv,Coelho:2010fv}. Lattice QCD provides a systematically improvable approach, but its application at finite baryon chemical potential is severely hindered by the sign problem. Furthermore, while the characteristic energy scales in the cores of compact stars are large, they remain in the vicinity of the hadronic scale and do not fully reach the asymptotic regime where pQCD becomes entirely reliable. Consequently, nonperturbative strong-interaction effects remain paramount \cite{Lattimer:2004pg,Song:2019qoh,Xia:2025waz}. The physics of compact stars thus remains an open problem, primarily limited by the lack of a reliable, continuous, and Poincar\'e-covariant nonperturbative framework for QCD at finite density.

The Dyson-Schwinger equations (DSEs) offer such a framework. As the fundamental equations of motion for QCD's Green functions, DSEs provide a continuum, symmetry-preserving approach capable of simultaneously addressing confinement and dynamical chiral symmetry breaking. Because this approach has achieved considerable success in describing hadron properties in the vacuum, it is highly promising for investigating the dense matter inside compact stars. To make the finite-density problem computationally tractable while rigorously retaining these essential nonperturbative features, we employ a symmetry-preserving vector$\,\otimes\,$vector contact interaction model within the DSE framework~\cite{gutierrez2010pion}. This approach is algebraically simple yet successfully preserves the essential global symmetries and dynamical features of QCD~ \cite{Ahmad:2016iez,Zhang:2020ecj,Xu:2021iwv,Yin:2021uom,Cheng:2022jxe,Zamora:2023fgl,Cheng:2024gyv,Paredes-Torres:2024mnz,Xing:2025eip}. 
In this work, we extend the quark gap equation with the contact interaction to the regime of zero temperature $T$ and finite quark chemical potential $\mu_f$ $(f=u,d,s)$. By numerically solving related equations, we obtain the dressed-quark propagators, which are inherently Poincar\'e-covariant and serve as the fundamental inputs for constructing the equation of state. We then systematically investigate the mass-radius profiles and tidal deformabilities of the resulting strange quark stars.

This paper is organized as follows. In Sec.~\ref{sec:quarkdse}, we introduce the general form of the gap equation at finite temperature and density, detailing the proper-time regularization scheme and its zero-temperature limit. In Sec.~\ref{sec:TOVsec}, we outline the thermodynamic framework for constructing the EOS, solving the TOV equations, and calculating the tidal deformability. In Sec.~\ref{sec:result}, we present our numerical results. We first vary the interaction coupling constant and the vacuum pressure, evaluating their impacts on the mass-radius relations, then analyze the dependence of the tidal deformability on the stellar mass, with a specific focus on canonical $1.4\,M_\odot$ stars. Furthermore, we extend our analysis to investigate the structural effects induced by changing the ultraviolet cutoff. Finally, a summary of our findings and concluding remarks is provided in Sec.~\ref{sec:summary}.

\section{QUARK gap EQUATION}
\label{sec:quarkdse}

To investigate the effects of finite $T$ and $\mu_f$ on the properties of strange quark stars, we first analyze the solutions of the quark gap equation. In Euclidean space, the inverse of the dressed-quark propagator takes the form \cite{Roberts:2000aa,Fischer:2011mz,Qin:2010nq,Li:2025cll},
\begin{equation}
    S^{-1}(\vec{p},\widetilde{\omega}_n) = i\vec{\gamma}\cdot \vec{p} + i\gamma_4\widetilde{\omega}_n + m + \Sigma(\vec{p},\widetilde{\omega}_n), 
    \label{eq:quarkgap1}
\end{equation}
with the quark self-energy given by
\begin{align}
     \Sigma(\vec{p},\widetilde{\omega}_n) = \int_{l,q} &g^2 D_{\mu\nu}(\vec{p}-\vec{q},\Omega_{nl};T,\mu_f)\gamma_\mu\frac{\lambda^a}{2} \nonumber \\
     &\times S(\vec{q},\widetilde{\omega}_l)\Gamma_\nu^a(\vec{p},\widetilde{\omega}_n,\vec{q},\widetilde{\omega}_l;T,\mu_f).
     \label{eq:selfenergy}
\end{align}
Here, $\widetilde{\omega}_n = \omega_n + i\mu_f$, where $\omega_n = (2n+1)\pi T$ are the fermionic Matsubara frequencies ($n \in \mathbb{Z}$), and $m$ is the current-quark mass. The integration symbol compactly denotes the Matsubara sum and three-momentum integration, $\int_{l,q} \equiv T\sum_{l=-\infty}^{\infty}\int \frac{d^3q}{(2\pi)^3}$. Furthermore, $D_{\mu\nu}$ is the dressed-gluon propagator, with $\Omega_{nl} = \omega_n - \omega_l = 2(n-l)\pi T$ being the bosonic Matsubara frequencies, and $\Gamma_\nu^a$ is the dressed-quark-gluon vertex.

In this work, we employ the rainbow-ladder (RL) truncation, $\Gamma_\nu^a = \frac{\lambda^a}{2}\gamma_{\nu}$ \cite{Roberts:2000aa,Maris:2003vk}. For the gluon propagator, we adopt a symmetry-preserving contact interaction. This momentum-independent model has been extensively and successfully applied to describe hadron properties~ \cite{Ahmad:2016iez,Zhang:2020ecj,Xu:2021iwv,Yin:2021uom,Cheng:2022jxe,Zamora:2023fgl,Cheng:2024gyv,Paredes-Torres:2024mnz,Xing:2025eip,Gutierrez-Guerrero:2026rsb},
\begin{equation}
    g^2D_{\mu\nu} = \delta_{\mu\nu}\frac{4\pi\alpha_{ir}}{m_G^2},
    \label{eq:contact}
\end{equation}
where $m_G$ is a gluon mass scale and $\alpha_{ir}$ represents the infrared value of the running coupling~\cite{Cui:2019dwv,dEnterria:2022hzv,Deur:2022msf}. Implementing this interaction within the rainbow truncation simplifies the gap equation to:
\begin{equation}
    S^{-1}(\vec{p},\widetilde{\omega}_n) = i\vec{\gamma}\cdot \vec{p} + i\gamma_4\widetilde{\omega}_n + m + \frac{16\pi}{3}\frac{\alpha_{ir}}{m_G^2}\int_{l,q}\gamma_{\mu} S(\vec{q},\widetilde{\omega}_l)\gamma_{\mu}.
    \label{eq:quarkgap2}
\end{equation}

The inverse of the dressed-quark propagator can be decomposed into its general Dirac structure \cite{Qin:2010nq}:
\begin{equation}
    S^{-1}(\vec{p},\widetilde{\omega}_n) = i\vec{\gamma}\cdot \vec{p} A(\vec{p},\widetilde{\omega}_n) + i\gamma_4\widetilde{\omega}_n C(\vec{p},\widetilde{\omega}_n) + B(\vec{p},\widetilde{\omega}_n). 
    \label{eq:propagator_ansatz}
\end{equation}
Inserting Eq.~\eqref{eq:propagator_ansatz} into Eq.~\eqref{eq:quarkgap2} and projecting out the scalar components via appropriate Dirac traces yields explicit integral equations for $A$, $B$, and $C$. Because the contact interaction is independent of the external momentum, these scalar functions must also be momentum-independent. Consequently, for any nonzero solution, they can be parameterized as,
\begin{align}
    A(\vec{p},\widetilde{\omega}_n) &= 1, \nonumber \\
    B(\vec{p},\widetilde{\omega}_n) &= M_f, \nonumber \\
    \widetilde{\omega}_n \left[C(\vec{p},\widetilde{\omega}_n)-1\right] &= -i \eta_f,
    \label{eq:defineRBRC}
\end{align}
where the constants $M_f$ and $\eta_f$ are determined by the coupled equations:
\begin{align}
    \eta_f &= \frac{8}{3\pi}\frac{\alpha_{ir}}{m_G^2}\int_{l,s} s^{\frac{1}{2}}\frac{ i\widetilde{\omega}_l'}{s+\widetilde{\omega}_l'^2+ M_f^2}, \nonumber \\
    M_f &= m + \frac{16}{3\pi}\frac{\alpha_{ir}}{m_G^2}\int_{l,s} s^{\frac{1}{2}}\frac{ M_f}{s+\widetilde{\omega}_l'^2+ M_f^2}.
    \label{eq:RBRC}
\end{align}
$M_f$ is the dressed quark mass. Here, we have defined an effective quark chemical potential $\mu'_f = \mu_f - \eta_f$ and the shifted frequency $\widetilde{\omega}_l' = \omega_l + i\mu'_f$, with $s = |\vec{q}\,|^2$. The integration measure is defined as $\int_{l,s} \equiv T\sum_{l=-\infty}^{\infty}\int_0^\infty ds$. Physically, the self-energy contribution $\eta_f$ acts as a dynamical shift to the quark chemical potential, meaning the effective quark chemical potential $\mu'_f$ experienced by the quarks is modified by interaction effects.

To render the integrals finite while preserving the relevant spacetime symmetries, we employ a proper-time regularization scheme. The momentum denominators are regulated as follows:
\begin{equation}
    \frac{1}{s+\widetilde{\omega}_l'^2+ M_f^2} \rightarrow \frac{1}{s+\widetilde{\omega}_l'^2+ M_f^2} \, e^{-x \left|s+\widetilde{\omega}_l'^2+ M_f^2\right|} \Bigg|_{x=\tau_{ir}^2}^{\tau_{uv}^2},
    \label{eq:PT}
\end{equation}
where $\tau_{ir} = 1/\Lambda_{ir}$ and $\tau_{uv} = 1/\Lambda_{uv}$. Here, $\Lambda_{uv}$ is the ultraviolet cutoff introduced to regularize divergent integrals and reflects the characteristic energy scale of the theory, while $\Lambda_{ir}$ is an infrared cutoff utilized to effectively simulate quark confinement. 

This framework allows us to solve the propagator numerically for arbitrary temperature and quark chemical potential. Since our primary focus is on cold, dense quark stars, we evaluate the system in the zero-temperature limit ($T \to 0$), which requires the discrete Matsubara sum to be replaced by a continuous energy integral:
\begin{equation}
    \lim_{T \to 0} T \sum_{l=-\infty}^{\infty} \longrightarrow \frac{1}{2\pi} \int_{-\infty}^{\infty} d\omega_l.
    \label{eq:Tiszero}
\end{equation}

\section{STRUCTURE OF STRANGE QUARK STARS}
\label{sec:TOVsec}
In the $T \to 0$ regime, our primary focus is the dependence of macroscopic physical quantities—such as quark number density, pressure, and energy density—on the quark chemical potential. Based on the quark propagator obtained in the previous section, the flavor-specific quark number density at zero temperature is given by \cite{Nickel:2006vf}
\begin{equation}
    n_f(\mu_f) = -N_c \int_{l,q} \mathrm{Tr}_{\mathrm{D}} \left\{\gamma_4 S(\vec{q},\widetilde{\omega}_l)\right\},
    \label{eq:numberdensity1}
\end{equation}
where $N_c = 3$ denotes the number of colors, and the trace $\mathrm{Tr}_{\mathrm{D}}$ is taken over Dirac indices. 

Evaluating this expression with our model, we obtain the explicit form:
\begin{equation}
\begin{split}
      n_f(\mu_f) &= N_c \frac{3}{8\pi}\frac{m_G^2}{\alpha_{ir}} \eta_f \\
     &\overset{T \to 0}{=} 
     \begin{cases}
        0, & \mu'_f < M_f, \\
        N_c \frac{\left(\mu_f'^{\,2} - M_f^2\right)^{\frac{3}{2}}}{3\pi^2}, & \mu'_f > M_f.
     \end{cases}
     \label{eq:numberdensity2}
\end{split} 
\end{equation}
This result clearly shows that the quark number density becomes nonzero only when the dynamically corrected quark chemical potential $\mu'_f$ exceeds the scalar piece $M_f$. Because $M_f$ is intimately related to the dynamical quark mass, this threshold behavior represents the onset of the Fermi sea, consistent with findings in Ref.~\cite{Lugones:2022upj}.

To determine the particle composition of dense matter within strange quark stars, we impose chemical equilibrium ($\beta$-equilibrium). The constituent particles achieve equilibrium via the weak interaction processes: $d \leftrightarrow u+e^-+\bar{\nu}_e$, $s \leftrightarrow u+e^-+\bar{\nu}_e$, and $u+d \leftrightarrow u+s$. Because neutrinos are assumed to escape freely from cold compact stars, their chemical potentials are set to zero ($\mu_{\nu} = 0$), and their contributions to the thermodynamics are neglected. Consequently, the chemical potentials of the species satisfy:
\begin{equation}
    \mu_s = \mu_d = \mu_u + \mu_e.
    \label{eq:equilibrium}
\end{equation}
This allows us to introduce two independent chemical potentials: the baryon chemical potential $\mu_B$ and the electric charge chemical potential $\mu_Q$. The individual quark chemical potentials are then parameterized as:
\begin{equation}
\begin{split}
    \mu_u &= \frac{1}{3}\mu_B + \frac{2}{3}\mu_Q, \\
    \mu_d &= \frac{1}{3}\mu_B - \frac{1}{3}\mu_Q, \\
    \mu_s &= \frac{1}{3}\mu_B - \frac{1}{3}\mu_Q, \\
    \mu_e &= -\mu_Q.
    \label{eq:baryonpotential}
\end{split}   
\end{equation}
Additionally, the stellar matter must satisfy local electric charge neutrality:
\begin{equation}
    \frac{2}{3}n_u - \frac{1}{3}n_d - \frac{1}{3}n_s - n_e = 0.
    \label{eq:chargeneutrality}
\end{equation}
By solving the charge neutrality condition Eq.~\eqref{eq:chargeneutrality} alongside the density relations, the electric chemical potential $\mu_Q$ is uniquely determined for any given $\mu_B$. 

Once the densities are known, the thermodynamic pressure for a given flavor can be computed by integrating the number density \cite{Zong:2008sm},
\begin{equation}
    P_f(\mu_f) = P_f(\mu_f)\big|_{\mu_f=0} + \int_{0}^{\mu_f} n_f(\tilde{\mu}) \, \mathrm{d}\tilde{\mu}.
    \label{eq:pressure}
\end{equation}
The integration constant represents the vacuum pressure at $\mu=0$. In the context of our treatment, it naturally plays the role of a quark confinement ``bag'' constant. Following the convention of the MIT bag model, we define the total vacuum pressure contribution as $-B$ ($B>0$) \cite{Wadhwa:2025mae}. The total pressure of the system is then given by:
\begin{equation}
    P = -B + \sum_{i=u,d,s,e} P_i(\mu_i).
    \label{eq:totalpressure}
\end{equation}
In the following sections, we will vary the value of $B$ to illustrate its effect on the stellar structure. Finally, standard thermodynamic relations yield the total energy density:
\begin{equation}
    \epsilon = -P + \sum_{i=u,d,s,e} \mu_i n_i(\mu_i),
    \label{eq:energydensity}
\end{equation}
which, together with Eq.~\eqref{eq:totalpressure}, provides the equation of state (EOS), $P(\epsilon)$.

With the EOS determined, we compute the macroscopic structure of strange quark stars by solving the Tolman-Oppenheimer-Volkoff (TOV) equations. Assuming spherical symmetry and using natural units ($G=c=1$), the equations are \cite{Tolman:1939jz,Oppenheimer:1939ne}:
\begin{align}
    \frac{\mathrm{d}P(r)}{\mathrm{d}r} &= -\frac{\left[\epsilon(r) + P(r)\right]\left[M(r) + 4\pi r^3 P(r)\right]}{r\left[r - 2M(r)\right]}, \nonumber \\
    \frac{\mathrm{d}M(r)}{\mathrm{d}r} &= 4\pi r^2 \epsilon(r),
    \label{eq:TOV}
\end{align}
where $M(r)$ is the gravitational mass enclosed within radius $r$. Integrating these equations from the center ($r=0$) to the surface ($r=R$, where $P(R)=0$) yields the mass-radius ($M-R$) relation ($M$ and $R$ are the star's total mass and radius), which can be compared against astrophysical measurements from pulsars such as PSR J0740+6620 \cite{Miller:2021qha,Riley:2021pdl}, PSR J0030+0451 \cite{Miller:2019cac,Riley:2019yda}, PSR J0437-4715 \cite{Choudhury:2024xbk,Miller:2025qfq}, and the central compact object in HESS J1731-347~\cite{Doroshenko:2022nwp}.

In addition to the mass and radius, gravitational-wave observations (e.g., GW170817) provide stringent constraints on the tidal deformability of compact stars. The dimensionless tidal deformability $\Lambda$ characterizes a star's response to an external tidal field and is highly sensitive to the underlying EOS. It is defined as \cite{Thorne:1997kt,Hinderer:2007mb,Hinderer:2009ca},
\begin{equation}
    \Lambda = \frac{2}{3}k_2 \left(\frac{R}{M}\right)^5,
    \label{eq:Lambda}
\end{equation}
where $k_2$ is the second-order tidal Love number, that is extracted by simultaneously solving the TOV equations and the following coupled differential equations for the metric perturbation $H(r)$ and its derivative $\beta(r)$,
\begin{align}
    \frac{\mathrm{d}H(r)}{\mathrm{d}r} &= \beta, \nonumber \\
    \frac{\mathrm{d}\beta(r)}{\mathrm{d}r} &= 2H\left(1 - \frac{2M}{r}\right)^{-1} \Bigg\{ -2\pi\left[5\epsilon + 9P + \frac{\mathrm{d}\epsilon}{\mathrm{d}P}(\epsilon + P)\right] \nonumber \\
    &\quad + \frac{3}{r^2} + 2\left(1 - \frac{2M}{r}\right)^{-1} \left(\frac{M}{r^2} + 4\pi r P\right)^2 \Bigg\} \nonumber \\
    &\quad + \frac{2\beta}{r}\left(1 - \frac{2M}{r}\right)^{-1} \left[ -1 + \frac{M}{r} + 2\pi r^2 (\epsilon - P) \right].
    \label{eq:tidal}
\end{align}
These equations are integrated outwards starting from the stellar center with the boundary conditions $H(r) \approx a_0 r^2$ and $\beta(r) \approx 2a_0 r$ as $r \to 0$. The arbitrary scaling constant $a_0$ cancels out in the final calculation of the Love number; for simplicity, we set $a_0 = 1$.

At the stellar surface $r=R$, the interior solution is matched to the exterior metric. The Love number $k_2$ is then evaluated as,
\begin{align}
    k_2 &= \frac{8C^5}{5}(1-2C)^2\bigl[2 + 2C(y-1) - y\bigr] \nonumber \\
    &\quad \times \biggl\{ 2C\bigl[6 - 3y + 3C(5y - 8)\bigr] \nonumber \\
    &\quad\quad + 4C^3\bigl[13 - 11y + C(3y - 2) + 2C^2(1 + y)\bigr] \nonumber \\
    &\quad\quad + 3(1-2C)^2\bigl[2 - y + 2C(y-1)\bigr]\ln(1-2C)\biggr\}^{-1},
    \label{eq:LOVEk2}
\end{align}
with the stellar compactness $C = M/R$ and the surface matching parameter defined as:
\begin{equation}
    y = \frac{R\beta(R)}{H(R)} - \frac{4\pi R^3 \epsilon_0}{M}.
    \label{eq:y_def}
\end{equation}
Crucially, for bare strange quark stars, the energy density does not vanish at the surface. The term $\epsilon_0$ represents this non-zero surface energy density, which introduces a finite discontinuity that must be accounted for in the evaluation of $y$.
\section{RESULTS AND DISCUSSION}\label{sec:result}

In this section, we discuss the influence of parameter choices. We begin with the parameters calibrated to reproduce hadronic properties in the vacuum ($T=0$, $\mu_f=0$) and subsequently introduce adjusted parameter schemes to describe strange quark stars at finite quark chemical potential. 

As the quark and baryon chemical potential becomes nonzero, the constituent quark density increases. Driven by the asymptotic freedom of QCD and in-medium screening effects, it is natural to expect a reduction in the effective coupling strength at high densities. To investigate this, we systematically reduce the infrared value of the coupling constant $\alpha_{ir}$ to examine its impact on the EOS and the macroscopic properties of quark stars. The parameter sets employed in our calculations are listed in Table~\ref{tab:couplingpara}. The first row corresponds to the vacuum parameters fitted to hadronic bound states, while the subsequent rows represent the dense-matter schemes with reduced coupling.

\begin{table}
\centering
\caption{\label{tab:couplingpara}
Parameter sets used in the contact interaction model. The first row corresponds to the parameters fitted to meson properties in the vacuum ($T=0$, $\mu_f=0$), which we refer to as the hadronic parameters. The subsequent sets are generated by reducing the infrared value of the coupling constant, $\alpha_{ir}$, while keeping the current-quark masses and cutoff scales fixed. Dimensioned quantities are in GeV.}
\setlength{\tabcolsep}{8pt}
\renewcommand{\arraystretch}{1.3}
\begin{tabular}{ccccc}
\toprule
$m_{u,d}$ & $m_s$ & $\alpha_{ir}$ & $\Lambda_{ir}$ & $\Lambda_{uv}$ \\
\colrule
0.007 & 0.170 & $0.9300\pi$ & 0.240 & 0.905 \\
0.007 & 0.170 & $0.8325\pi$ & 0.240 & 0.905 \\
0.007 & 0.170 & $0.7350\pi$ & 0.240 & 0.905 \\
\botrule
\end{tabular}
\end{table}

\begin{figure}
\centering
\includegraphics[width=0.9\columnwidth]{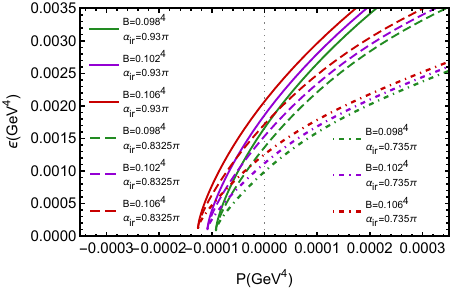} 
\caption{\label{fig:EOScoupling} (Color online) The calculated equations of state, plotted as energy density $\epsilon$ versus pressure $P$. The red, purple, and green curves correspond to bag parameters $B=(0.106\,\mathrm{GeV})^4$, $(0.102\,\mathrm{GeV})^4$, and $(0.098\,\mathrm{GeV})^4$, respectively. Solid lines represent the infrared value of the running coupling $\alpha_{ir}=0.9300\pi$, dashed lines correspond to $\alpha_{ir}=0.8325\pi$, and dot-dashed lines to $\alpha_{ir}=0.7350\pi$.} 
\end{figure}

The resulting equations of state are presented in Fig.~\ref{fig:EOScoupling}. We calculate the EOS for the three different values of $\alpha_{ir}$ and pair each with three distinct vacuum bag pressures: $B=(0.098\,\mathrm{GeV})^4$, $(0.102\,\mathrm{GeV})^4$, and $(0.106\,\mathrm{GeV})^4$. The comparison reveals two clear trends. First, for a given coupling strength $\alpha_{ir}$, a smaller bag parameter $B$ yields a stiffer EOS. Second, for a fixed $B$, a reduction in $\alpha_{ir}$ also leads to a stiffer EOS. These behaviors align with physical expectations: a lower vacuum energy penalty or a weaker effective interaction reduces the overall softening of the EOS, leading to a more rapid increase of pressure with respect to energy density. As is well established in the study of compact stars, the stiffness of quark matter is highly sensitive to the interplay between the vacuum pressure and the interaction strength \cite{Li:2021teo,Li:2019akk}.

\begin{figure}
\centering
\includegraphics[width=0.9\columnwidth]{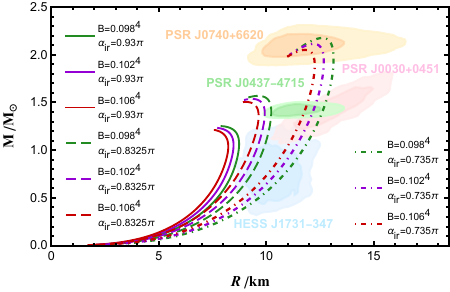} 
\caption{\label{fig:TOVcoupling} (Color online) Mass-radius relations of strange quark stars obtained for different values of $B$ and $\alpha_{ir}$ (line styles match those in Fig.~\ref{fig:EOScoupling}). The shaded regions represent observational constraints from various pulsars and the central compact object in HESS J1731-347. The darker central areas indicate the $68\%$ confidence intervals, while the lighter surrounding areas represent the $95\%$ confidence intervals.} 
\end{figure}

To evaluate the viability of these EOS models, we integrate the TOV equations to obtain the $M-R$ relations for strange quark stars. The results are compared with multi-messenger astronomical constraints in Fig.~\ref{fig:TOVcoupling}, including observational data from PSR J0740+6620 \cite{Miller:2021qha,Riley:2021pdl}, PSR J0030+0451 \cite{Miller:2019cac,Riley:2019yda}, PSR J0437-4715 \cite{Choudhury:2024xbk,Miller:2025qfq}, and HESS J1731-347 \cite{Doroshenko:2022nwp}. 

As shown by the solid curves in Fig.~\ref{fig:TOVcoupling}, the EOS derived using the pure hadronic vacuum parameters ($\alpha_{ir}=0.9300\pi$) is exceedingly soft, yielding a maximum stellar mass of only $\sim 1.2\,M_\odot$. Even when $B$ is minimized, the curve barely touches the lower boundary of the HESS J1731-347 mass-radius region. This demonstrates that vacuum parameters are fundamentally inadequate for describing cold, dense strange quark stars.

Conversely, as the coupling constant is reduced to simulate in-medium effects, the EOS stiffens, allowing the star to support a larger maximum mass. For $\alpha_{ir}=0.8325\pi$, the $M-R$ curves begin to satisfy the constraints from PSR J0437-4715 and HESS J1731-347. Notably, when the coupling is further reduced to $\alpha_{ir}=0.7350\pi$, the mass-radius profiles fall into excellent agreement with the majority of the current observational regions, including the massive pulsar PSR J0740+6620. This adjustment validates the necessity of treating the effective coupling dynamically in dense environments and serves as a robust starting point for modeling compact stars within a Poincaré-covariant framework.

\begin{figure}
\centering
\includegraphics[width=0.9\columnwidth]{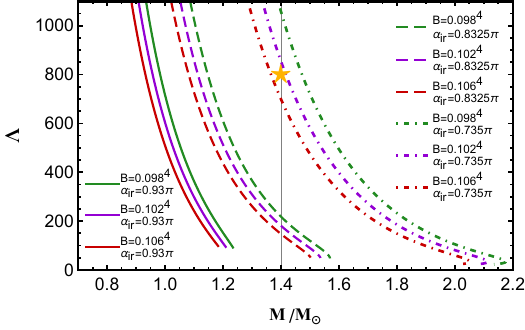} 
\caption{\label{fig:LOVEcoupling} (Color online) The dimensionless tidal deformability $\Lambda$ as a function of stellar mass $M$ for different parameter sets (line styles and colors match those in Fig.~\ref{fig:EOScoupling}). The yellow star marks the value of $\Lambda$ for a canonical $1.4\,M_\odot$ star.} 
\end{figure}

In Fig.~\ref{fig:LOVEcoupling}, we plot the dimensionless tidal deformability $\Lambda$ as a function of stellar mass. As expected, $\Lambda$ decreases monotonically with increasing stellar mass, reflecting the fact that more massive, highly compact stars are less susceptible to external tidal deformations. Because the $\alpha_{ir}=0.9300\pi$ set fails to reach a maximum mass of $1.4\,M_\odot$, it is excluded from tidal deformability analyses related to GW170817. For the remaining two coupling values, the specific properties of a canonical $1.4\,M_\odot$ star are summarized in Table~\ref{tab:tidala735}.

\begin{table}[htbp]
\centering
\caption{\label{tab:tidala735} Properties of a $1.4\,M_\odot$ strange quark star calculated using $\alpha_{ir}=0.7350\pi$ and $\alpha_{ir}=0.8325\pi$.}
\setlength{\tabcolsep}{10pt}
\renewcommand{\arraystretch}{1.3}
\begin{tabular}{ccccc}
\toprule
$\alpha_{ir}$ & B & $M/R$ & $k_2$ & $\Lambda_{1.4\,M_\odot}$ \\ 
\colrule
 & $(0.098\,\mathrm{GeV})^4$ & 0.1639 & 0.1881 & 1057.6 \\ 
$0.7350\pi$ & $(0.102\,\mathrm{GeV})^4$ & 0.1699 & 0.1816 & 854.1 \\ 
 & $(0.106\,\mathrm{GeV})^4$ & 0.1757 & 0.1753 & 699.1 \\ 
\colrule
 & $(0.098\,\mathrm{GeV})^4$ & 0.2020 & 0.1101 & 218.1 \\ 
$0.8325\pi$ & $(0.102\,\mathrm{GeV})^4$ & 0.2082 & 0.1061 & 180.9 \\ 
 & $(0.106\,\mathrm{GeV})^4$ & 0.2146 & 0.1007 & 147.3 \\ 
\botrule
\end{tabular}
\end{table}

Based on these results, we find that the combination of $\alpha_{ir}=0.7350\pi$ and $B = (0.106\,\mathrm{GeV})^4$ provides an optimal description of strange quark stars. Not only does this parameter set satisfy the strict mass-radius constraints from multiple pulsars, but it also yields a tidal deformability ($\Lambda_{1.4\,M_\odot} \approx 699$) that is entirely consistent with the upper limits extracted from the GW170817 event.

Finally, we consider the effect of the characteristic energy scale. As the system transitions from vacuum to the extreme densities found in compact stars, the relevant energy scale increases. Consequently, it is physically well-motivated to shift the ultraviolet cutoff $\Lambda_{uv}$ to larger values. To investigate this, we fix the vacuum pressure at $B = (0.106\,\mathrm{GeV})^4$ and the current-quark masses, while varying $\Lambda_{uv}$ and $\alpha_{ir}$. We increase the UV cutoff by $10\%$ relative to the vacuum value, exploring the four parameter combinations detailed in Table~\ref{tab:uvandcouplepara}. 

\begin{table}[htbp]
\centering
\caption{\label{tab:uvandcouplepara}
Parameter sets used to study the effect of the UV cutoff at a fixed bag pressure $B=(0.106\,\mathrm{GeV})^4$. We compare the baseline dense-matter set (Row 1) with three modifications: a $10\%$ increase in $\Lambda_{uv}$ (Row 2), a $20\%$ reduction in $\alpha_{ir}$ (Row 3), and a simultaneous adjustment of both (Row 4). Dimensioned quantities are in GeV.}
\setlength{\tabcolsep}{8pt}
\renewcommand{\arraystretch}{1.3}
\begin{tabular}{ccccc}
\toprule
$m_{u,d}\ $ & $m_s\ $ & $\alpha_{ir}$ & $\Lambda_{ir}\ $ & $\Lambda_{uv}\ $ \\
\colrule
0.007 & 0.170 & $0.7350\pi$ & 0.240 & 0.9050 \\
0.007 & 0.170 & $0.7350\pi$ & 0.240 & 0.9955 \\
0.007 & 0.170 & $0.5880\pi$ & 0.240 & 0.9050 \\
0.007 & 0.170 & $0.5880\pi$ & 0.240 & 0.9955 \\
\botrule
\end{tabular}
\end{table}


\begin{figure}[htbp]
\centering
\includegraphics[width=0.9\columnwidth]{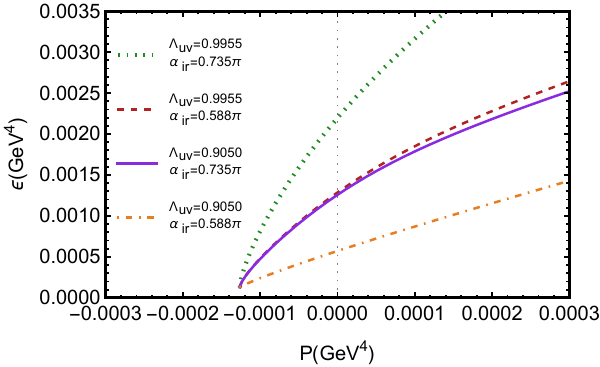} 
\caption{\label{fig:EOSuvcoupling} (Color online) Comparison of the EOSs using the four parameter combinations listed in Table~\ref{tab:uvandcouplepara}, with a fixed vacuum pressure $B = (0.106\,\mathrm{GeV})^4$.} 
\end{figure}

The corresponding equations of state are shown in Fig.~\ref{fig:EOSuvcoupling}. Consistent with our previous findings, comparing the $\alpha_{ir}=0.7350\pi$ curve with the $\alpha_{ir}=0.5880\pi$ curve confirms that decreasing the coupling constant stiffens the EOS. However, for a fixed coupling strength, increasing the UV cutoff from $0.9050\,\mathrm{GeV}$ to $0.9955\,\mathrm{GeV}$ results in a softer EOS. 

\begin{figure}[htbp]
\centering
\includegraphics[width=0.9\columnwidth]{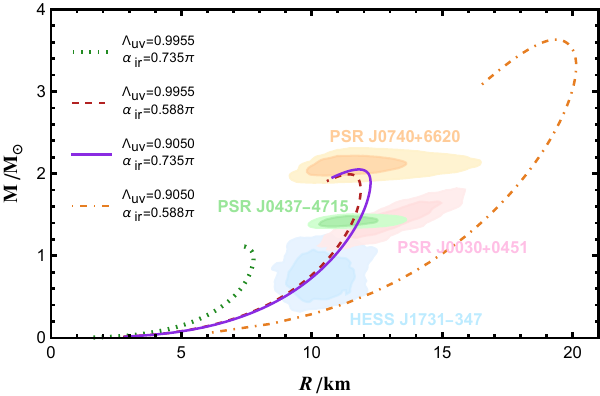} 
\caption{\label{fig:TOVuvcoupling} (Color online) Mass-radius relations obtained using the four parameter combinations listed in Table~\ref{tab:uvandcouplepara}, with a fixed vacuum pressure $B = (0.106\,\mathrm{GeV})^4$.} 
\end{figure}

This softening behavior has immediate consequences for the stellar structure, as illustrated by the TOV solutions in Fig.~\ref{fig:TOVuvcoupling}. Increasing the UV cutoff alone (green dotted curve) drastically softens the EOS, yielding a maximum stellar mass of only $1.12\,M_\odot$, which is unphysical. Conversely, both the baseline dense-matter set (purple solid curve: $\alpha_{ir}= 0.7350\pi$, $\Lambda_{uv} = 0.9050\,\mathrm{GeV}$) and the simultaneously adjusted set (red dashed curve: $\alpha_{ir} = 0.5880\pi$, $\Lambda_{uv}= 0.9955\,\mathrm{GeV}$) provide excellent descriptions of current observational data. This demonstrates that within our framework, shifting the UV cutoff to higher energy scales mandates a simultaneous reduction in the coupling constant to maintain phenomenological consistency. This intricate balance directly mirrors the running of the QCD coupling, where higher energy scales are associated with weaker effective interactions, consistent with the findings in Ref.~\cite{Yin:2019bxe}.

\begin{figure}[htbp]
\centering
\includegraphics[width=0.9\columnwidth]{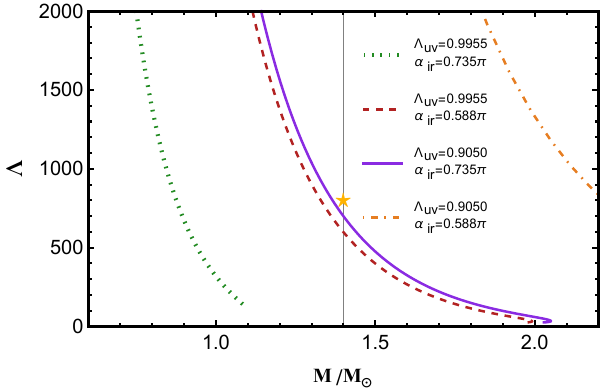} 
\caption{\label{fig:LOVEuvcoupling} (Color online) Tidal deformability curves corresponding to the four parameter combinations listed in Table~\ref{tab:uvandcouplepara}, with a fixed vacuum pressure $B = (0.106\,\mathrm{GeV})^4$.} 
\end{figure}

Finally, Fig.~\ref{fig:LOVEuvcoupling} displays the tidal deformabilities for these varied UV cutoff scenarios. Because the case with an increased $\Lambda_{uv}$ alone (green dotted curve) cannot support a $1.4\,M_\odot$ star, its canonical tidal deformability is undefined. On the other hand, reducing only the coupling constant (orange dot-dashed curve) over-stiffens the EOS, yielding an excessively large tidal deformability of $\Lambda_{1.4\,M_\odot} \approx 6600$, which violates current gravitational-wave constraints. However, applying the simultaneous adjustment of increasing $\Lambda_{uv}$ and decreasing $\alpha_{ir}$ (red dashed curve) strikes the proper balance, predicting a reasonable tidal deformability of $\Lambda_{1.4\,M_\odot} \approx 598$.
\section{SUMMARY AND CONCLUSIONS}\label{sec:summary}

In this study, we investigated the structural and thermodynamic properties of strange quark stars using a symmetry-preserving vector$\,\otimes\,$vector contact interaction model within the Dyson-Schwinger equation framework at finite quark chemical potential. By computing the Poincar\'e-covariant quark propagator at zero temperature, we derived the equation of state (EOS) for dense strange quark matter, and systematically explored its astrophysical implications through the dependence on key model parameters.

Our results demonstrate that the stiffness of the EOS is highly sensitive to the effective coupling strength and the ultraviolet energy scale. Specifically, we found that decreasing the coupling constant leads to a stiffer EOS, while increasing the ultraviolet cutoff results in a softer one. Driven by the physical expectation of asymptotic freedom and in-medium screening, we adjusted these parameters to account for the dense stellar environment. We identified two optimal parameter sets—$\alpha_{ir} = 0.735\pi$, $\Lambda_{uv} = 0.905\,\mathrm{GeV}$, and $\alpha_{ir} = 0.588\pi$, $\Lambda_{uv} = 0.9955\,\mathrm{GeV}$—which yield mass-radius relations and tidal deformabilities in excellent agreement with current multi-messenger observational data. These agreements hold even when the vacuum bag pressure $B$ is moderately varied around $(0.106\,\mathrm{GeV})^4$. 

Crucially, our analysis highlights the necessity of dynamically adjusting the effective coupling when shifting the ultraviolet cutoff to higher energy scales, a behavior that inherently reflects the running of the strong coupling in quantum chromodynamics (QCD). These trends are consistently reflected in the macroscopic observables, including the maximum stellar masses and the dimensionless tidal deformabilities of $1.4\,M_\odot$ stars ($\Lambda_{1.4\,M_\odot}$). 

Overall, this work provides a comprehensive study of dense quark matter within a continuous, nonperturbative, and Poincar\'e-covariant framework, reaffirming the vital role of fundamental strong interactions in governing the macroscopic properties of compact stars. Future extensions of this work will aim to incorporate momentum-dependent interactions and more realistic QCD-inspired interaction kernels~\cite{Zhang:2026iuw} to further refine our understanding of the QCD phase diagram at extreme densities.

\begin{acknowledgements}
This work was supported by Natural Science Foundation of Jiangsu Province (grant no. BK20220122) and National Natural Science Foundation of China (grant no. 12233002). Use of the computing resources at \href{www.syli.cloud}{Jiangsu Xilixi Technology} is gratefully acknowledged.
\end{acknowledgements}

\bibliographystyle{apsrev4-1}
\bibliography{refs}



\end{document}